\documentstyle[twocolumn,pre,aps,epsf]{revtex}
\newcommand{\be}{\begin{equation}}
\newcommand{\ee}{\end{equation}}
\newcommand{\bear}{\begin{eqnarray}}
\newcommand{\eear}{\end{eqnarray}}

\def\Tc{{T_{\rm c}}}

\def\TMTSFX{{${\rm (TMTSF)_2X}$}}
\def\Ru{Sr$_2$RuO$_4$}
\def\Tc{{T_{\rm c}}}

\begin{document}
\draft

%% FOR TWO COLUMN  ACTIVATE THE LINE BELOW
\twocolumn[\hsize\textwidth\columnwidth\hsize\csname @twocolumnfalse\endcsname

%%%%%%%%%%%%%%%%%%%%%%%%%%%%%%%%%%%%%%%%%%%%%%%%%%%%%%%%%%%%%%%%%%%%%%%

\title{
Spin-triplet superconductivity
in quasi-one dimension}

\author{
Mahito Kohmoto and Masatoshi Sato}

\address{
Institute for Solid State Physics, University of Tokyo, 7-22-1
Roppongi, Minato-ku, Tokyo, Japan}

\maketitle
%\begin{center}
%\today
%\end{center}

\begin{abstract}
We consider a  system with electron-phonon interaction,
antiferromagnetic fluctuations and disconnected open Fermi
surfaces. The existence of odd-parity superconductivity in this
circumstance is shown for the first time. If it is applied to the
quasi-one-dimensional systems like the organic conductors \TMTSFX~ we
obtain spin-triplet superconductivity with nodeless gap. Our result is
also valid in higher dimensions(2d and 3d).
\end{abstract}


%% FOR TWO COLUMN ACTIVATE THE LINE BELOW
]

\narrowtext

In recent years the signs for unconventional superconductivity in many
compounds have been accumulated. Examples include high $\Tc$
superconductors, heavy fermions, organic conductors, \Ru, etc. The common
features of those compounds are quasi-low dimensionality and proximity of
antiferromagnetic(AF) order.

Here we consider seemingly the simplest
quasi-one-dimensional systems which are realized in (TMTSF)$_2$X family
where X$=$ PF$_6$, AsF$_6$, SbF$_6$, ClO$_4$, etc.(Bechgaard salts). At
ambient pressure, most of these extremely anisotropic compounds undergo a
metal-insulator transition at low temperature and have a
spin-density-wave(SDW) fundamental state. Under moderate pressure, the SDW
instability is suppressed and replaced by a superconducting transition
at a critical temperature of order of 1K
\cite{iy}. (One exception to this is (TMTSF)$_2$ClO$_4$ which is
superconducting at ambient pressure.) Thus these compounds may be
characterized by competition between superconducting and SDW ground
states \cite{jerome}.

As for the gap symmetry, Takigawa et al.\cite{takigawa} measured nuclear
relaxation rate of proton in (TMTSF)$_2 $ClO$_4$.
Their results show unconventional superconductivity(absence of coherence
peak) and the existence of nodes. However their measurement is restricted
to $T> 0.5\Tc$ and the existence of nodes are not yet conclusive. More
recently, Belin and Behnia \cite{bb} showed some evidences for nodeless gap
by thermal conductivity measurement. Also  experiments on
(TMTSF)$_2$ClO$_4$ and (TMTSF)$_2$PF$_6$ show that  $H_{\rm c2}$ for ${\vec
B}\parallel {\vec a}$ and ${\vec B}\parallel {\vec b}$ far exceed
the Pauli-limiting field $B_{\rm P} \sim \Delta _0 / ({\sqrt 2})
\mu_{\rm B})\sim 2$ Tesla for Bechgaard salts, where $\Delta_0$ is the 
superconducting order parameter at $T=0$ and $\mu_{\rm B}$ is the
Bohrmagneton\cite{chaikin1,chaikin2}. This suggests spin-triplet
superconductivity. Maki et al. studied theoretically impurity effects and
vortex states on these compounds \cite{maki}.

In this paper we consider AF magnetic coupling and electron-phonon
interaction under open disconnected Fermi surfaces.
The spin-triplet superconductivity is possible under  electron-phonon
interaction and AF magnetic coupling \cite{sk}.
We assume that AF fluctuation is not strong enough to give SDW
gap. We shall show the existence of spin-triplet
superconductivity in this circumstance for the first time. Our result
is also applicable to higher dimensions(2d and 3d).

{\it -- Open disconnected Fermi surface}\\
The Fermi surface of the system we consider is quasi-one-dimensional and
consists of two separated parts, which are $(k_{\rm F})_x\sim c>0$ and
$(k_{\rm F})_x\sim -c$.
($c$ is a constant.)
It is also supposed to be symmetric under parity transformation
$k\rightarrow -k$.
This type of Fermi surface is realized in \TMTSFX~.
The non-interacting part of the Hamiltonian of \TMTSFX~
is often written
\begin{equation}
E = -2t_a \cos(k_x a)  -2t_b \cos(k_y b).
\end{equation}
Here the ratio of $t_a$ and $t_b$ is about 10:1.

We take two interactions between electrons: one comes from
electron-phonon coupling, the other comes from AF
fluctuations.

{\it -- Phonon-mediated interaction} \\
It is written
\begin{equation}
H_{\rm int}^{\rm ph}
= - \sum_{kk'q}{ \sum_{\alpha \beta}f(q) { a_{k+q,\alpha}^\dagger
a_{k,\alpha} a_{k'-q,\beta}^\dagger a_{k',\beta}}},    \label{Hph}
\end{equation}
where $a$ and $a^\dagger$ are the usual fermion operators, $\alpha$ and
$\beta$ represent spin orientations and
$f(q) >0$ has a peak at $q = 0$.
We assume $f(q)=f(-q)$.
If one considers pairing interaction between
$k$ and $-k$, (\ref{Hph}) reduces to
\begin{equation}
H^{\rm ph}_{\rm int}
=-\frac{1}{2}\sum_{k,q}f(q)\left(\phi_0^{\dagger}(k+q)\phi_0(k)
+\vec{\phi}^{\dagger}(k+q)\cdot\vec{\phi}(k)\right),
\label{eqn:ph2}
\end{equation}
where
\begin{eqnarray}
\phi_0(k)=a_{k,\alpha}(\sigma_2)_{\alpha,\beta}a_{-k,\beta},\\
\vec{\phi}(k)=a_{k,\alpha}(\sigma_2\vec{\sigma})_{\alpha,\beta}a_{-k,\beta}.
\end{eqnarray}
Here $\phi_0(k)$ is the spin-singlet pairing and $\vec{\phi}(k)$
is the spin-triplet one:
\begin{equation}
a_{k,\alpha}a_{-k,\beta}=-\frac{1}{2}(\sigma_2)_{\alpha\beta}\phi_0(k)
+\frac{1}{2}(\vec{\sigma}\sigma_2)_{\alpha\beta}\cdot\vec{\phi}(k).
\end{equation}
Since $f(q)$ has a peak at $q=0$, this interaction is approximated by
\begin{eqnarray}
H^{\rm ph}_{\rm int}
&=&-\frac{1}{2}\sum_{k}f(0)\left(\phi_0^{\dagger}(k)\phi_0(k)
+\vec{\phi}^{\dagger}(k)\cdot\vec{\phi}(k)\right)
\nonumber\\
&=&-\sum_{k_x>0}f(0)\left(\phi_0^{\dagger}(k)\phi_0(k)
+\vec{\phi}^{\dagger}(k)\cdot\vec{\phi}(k)\right).
\label{eqn:ph3}
\end{eqnarray}

--{\it Interaction due to AF fluctuations} \\
It is written
\begin{eqnarray}
H_{\rm int}^{\rm AF}
= - \sum_{kk'q}{ \sum_{\alpha \beta \gamma \delta} {J(q)
{\vec{\sigma}_{\alpha \beta}}\cdot {\vec{\sigma}}_{\gamma \delta}
a_{k+q,
\alpha}^\dagger a_{k,\beta} a_{k'-q,\gamma}^\dagger a_{k',\delta}}},
\end{eqnarray}
where $J(q) > 0$ represents interaction due to AF fluctuations
and has a peak value at a nesting vector $q=\pm Q$.
The nesting vector $Q$ connects the two separated Fermi surfaces: $Q_x\sim 2c$.
We assume $J(q)=J(-q)$.
A similar analysis to above leads to
\begin{equation}
H^{\rm AF}_{\rm int}
=\frac{1}{2}\sum_{k,q}J(q)\left(3\phi_0^{\dagger}(k+q)\phi_0(k)
-\vec{\phi}^{\dagger}(k+q)\cdot\vec{\phi}(k)\right),
\end{equation}
and this interaction is approximated by
\begin{eqnarray}
H^{\rm AF}_{\rm int}
&&=\frac{1}{2}\sum_{k}J(Q)\left(3\phi_0^{\dagger}(k+Q)\phi_0(k)
-\vec{\phi}^{\dagger}(k+Q)\cdot\vec{\phi}(k)\right)
\nonumber\\
&&\hspace{-3ex}
+\frac{1}{2}\sum_{k}J(-Q)\left(3\phi_0^{\dagger}(k-Q)\phi_0(k)
-\vec{\phi}^{\dagger}(k-Q)\cdot\vec{\phi}(k)\right).
\nonumber\\
\end{eqnarray}
For fermions on the Fermi surface, this
becomes
\begin{eqnarray}
&&H^{\rm AF}_{\rm int}
\nonumber\\
&&=\frac{1}{2}\sum_{k_x\sim -c}J(Q)\left(3\phi_0^{\dagger}(k+Q)\phi_0(k)
-\vec{\phi}^{\dagger}(k+Q)\cdot\vec{\phi}(k)\right)
\nonumber\\
&&+\frac{1}{2}\sum_{k_x\sim c}J(-Q)\left(3\phi_0^{\dagger}(k-Q)\phi_0(k)
-\vec{\phi}^{\dagger}(k-Q)\cdot\vec{\phi}(k)\right)
\nonumber\\
&&=\sum_{k_x,k'_x\sim c}J(Q)\left(3\phi_0^{\dagger}(k')\phi_0(k)
+\vec{\phi}^{\dagger}(k')\cdot\vec{\phi}(k)\right).
\label{eqn:af}
\end{eqnarray}
Here $k'$ satisfies $k'=Q-k$.

--{\it Spin-triplet superconductivity}\\
Equation(\ref{eqn:ph3}) shows that the phonon-mediated interaction of
spin-triplet pairing has the same magnitude as that of spin-singlet one.
In many systems, however, only the spin-singlet superconductivity is
realized.
This is because the Fermi surfaces of usual matters are connected.
In a system of connected Fermi surface, the requirement of parity and
continuity does not admit the constant spin-triplet gap.
Therefore, the s-wave superconductivity is favored from the kinematic
reason.

The quasi-one-dimensional system we consider, however, has disconnected
Fermi surfaces, so it admits constant spin-triplet gap:
$\vec{d}|_{k_x\sim c}=-\vec{d}|_{k_x\sim -c}={\rm const.}$.
\footnote{$\vec{d}$ and $\psi$ is defined as
$\Delta(k)=i\sigma_2\psi(k)+i(\vec{d(k)}\cdot\vec{\sigma})\sigma_2$
\cite{leggett}.}
Therefore, there is no reason to prefer s-wave superconductivity.
{}From (\ref{eqn:af}), the interaction due to AF fluctuation disturbs the
spin-singlet superconductivity more than the spin-triplet one, so the
spin-triplet superconductivity is realized.
This spin-triplet gap is nodeless.

--{\it  BCS analysis}\\
Finally, we apply the BCS weak coupling theory. Since $\Tc$ of
(TMTSF)$_2$X is rather low, one can expect that the strong coupling
corrections do not change our results qualitatively. For the sake of
simplicity, the cutoffs of the phonon-mediated interaction and AF one are
taken to be same and denoted  by
$\hbar \omega_{\rm D}$.
We approximate the spin-singlet gap by
\begin{equation}
\psi |_{k_x\sim c}=\psi |_{k_x\sim -c}={\rm const.},
\end{equation}
and the spin-triplet gap by
\begin{equation}
\vec{d} |_{k_x\sim c}=-\vec{d} |_{k_x\sim -c}={\rm const.}.
\end{equation}
If we denote the density of states on the Fermi surface as $N(k_{\rm F})$,
the critical temperature of the spin-singlet superconductivity becomes
\cite{deGennes}
\begin{equation}
T_{\rm c}^{\rm even}
= 1.13 \hbar\omega_{\rm D}e^{1/\{N(k_{\rm F})V^{\rm even} \}},
\label{tceven}
\end{equation}
where
\begin{equation}
V^{\rm even}
= -f(0)+3 J(Q), \label{veven}
\end{equation}
and that of spin-triplet becomes
\begin{equation}
T_{\rm c}^{\rm odd}
= 1.13\hbar\omega_{\rm D}e^{1/\{N(k_{\rm F})V^{\rm odd}\}},
\label{tcodd}
\end{equation}
where
\begin{equation}
V^{\rm odd}
= -f(0)+J(Q). \label{vodd}
\end{equation}
{}From (\ref{veven}) and (\ref{vodd}),
the condition
$V^{\rm odd} < V^{\rm even}$
is always satisfied. Thus
$T_{\rm c}^{\rm odd} >T_{\rm c}^{\rm even}$ and
spin-triplet superconductivity is present if
$V^{\rm odd} = -f(0) + J(Q) <0$.
This simple result show that there is no superconductivity if AF
fluctuations dominate and we have spin-triplet superconductivity if
phonon-mediated interaction dominates.
Since $\vec{d}(k)$ is an odd function, the gaps at $(k_{\rm F})_x=\pm c$ have
opposite signs and there is no node in the gap.

%\vspace{\baselineskip}

%\noindent
%{\bf Acknowledgment} \\

%%%%%%%%%%%%%%%%%%%%%%%%%%%%%%%%%%%%%%%%%%%%%%%%%%%%%%%%%%%%%%%%%%%%%%%
%%  References                                                       %%
%%%%%%%%%%%%%%%%%%%%%%%%%%%%%%%%%%%%%%%%%%%%%%%%%%%%%%%%%%%%%%%%%%%%%%%

\end{document}